\documentclass[3p,times]{elsarticle}
\usepackage{ecrc}
\usepackage{epsfig}
\usepackage{epstopdf}
\def\bea#1\eea{\begin{align}#1\end{align}} 
\newcommand{\nnu}{\nonumber\\}
\newcommand{\bef}{\begin{figure}[htb]\centering}
\newcommand{\eef}{\end{figure}}

\volume{00}

\firstpage{1}

\journalname{Nuclear Physics A}

\runauth{Z. Kang et al.}

\jid{nupha}

\jnltitlelogo{Nuclear Physics A}

\usepackage{graphicx}
\usepackage{amsmath,amssymb}

\begin{document}

\begin{frontmatter}

\title{Transverse momentum broadening at NLO and QCD evolution of $\hat q$}

\author[LANL]{Hongxi Xing}
\author[LANL]{Zhong-Bo Kang}
\author[CCNU]{Enke Wang}
\author[CCNU,LBNL]{Xin-Nian Wang}

\address[LANL]{Theoretical Division, Los Alamos National Laboratory, Los Alamos, NM 87545, USA}
\address[CCNU]{Institute of Particle Physics and Key Laboratory of Lepton and Quark Physics (MOE), Central China Normal University, Wuhan 430079, China}
\address[LBNL]{Nuclear Science Division, Lawrence Berkeley National Laboratory, Berkeley, California 94720, USA}

\begin{abstract}
Within the framework of a high-twist approach, we show the first complete next-to-leading order calculation of transverse momentum broadening in semi-inclusive deeply inelastic $e+A$ scattering and Drell-Yan dilepton production in $p+A$ collisions. We demonstrate at one-loop level how QCD factorization holds for multiple scattering in nuclear medium, and the university of the associated quark-gluon correlation function. Our calculation also identifies QCD evolution equation for the quark-gluon correlation function, which determines the QCD factorization scale and energy dependence of jet transport parameter $\hat q$.  
\end{abstract}


\end{frontmatter}

\section{Introduction}
\label{intro}
The major goal in jet quenching physics is to extract the information of the medium created in heavy ion collisions, which is characterized by the jet transport parameter $\hat q$ \cite{Burke:2013yra}. So far, almost all the efforts in extracting the properties of the medium are based on perturbative QCD calculations of multiple parton interaction in the medium. However, it is not clear whether the properties of the medium such as the $\hat q$ as probed by a propagating jet is unique and intrinsic to the medium, independent of the hard processes that produce the energetic jet. This is a problem of factorization of multiple scattering in QCD \cite{Kang:2014lha} and so far has eluded many theoretical efforts. This is largely due to the complexity of QCD dynamics for multiple scattering, which involves both initial-state and final-state interactions and medium evolution. In this talk, we show the first complete next-to-leading order (NLO) calculation of transverse momentum broadening in semi-inclusive deeply inelastic $e+A$ scattering (SIDIS) and Drell-Yan dilepton production (DY) in $p+A$ collisions \cite{Kang:2013raa}. In these two processes, we have either virtual photon prepared in initial state or observed in final state, therefore one can always concentrate on one particular multiple scattering effect (either initial-state or final-state) to simplify the evaluation \cite{Wang:2001ifa, Xing:2011fb}, and make the analysis of QCD factorization for multiple scattering more clear.

\section{NLO transverse momentum broadening and QCD factorization in SIDIS}
%It is well known that through the measurements of the single inclusive hadron produced in $e+N$ scatterings, one can extract the fundamental structure of the nucleon. For example, the parton distribution function can be extracted through global analysis of the experimental data, where the analysis relies on the QCD factorization theorem at leading twsit (or single scattering), meaning the cross section can be factorized into the nonpeturbative parton distribution function and the perturbative calculable hard part. But when we go to $e+A$ collisions, multiple scattering becomes important and leads to nontrivial nuclear effect. In the parton model like picture, the nontrivial nuclear effect is closely related to the nonperturbative twist-4 parton correlation functions, which reprensent the fundamental structure of the nucleus. 

Transverse momentum broadening, $\Delta \langle\ell_{hT}^2\rangle =  \langle\ell_{hT}^2\rangle_{e+A} - \Delta \langle\ell_{hT}^2\rangle_{e+p}$, has long been recognized as one of the excellent probes to the QCD dynamics of multiple scattering and the medium properties as characterized by the twist-4 multi-parton correlation functions \cite{Guo:1998rd}.  
In SIDIS, because the virtual photon in initial state does not have strong interaction with the nuclear medium, the transverse momentum broadening for the produced hadron comes from final-state multiple scattering only, and the dominant contribution is from the double scattering,
\bea
\Delta \langle\ell_{hT}^2\rangle \approx \frac{d\langle \ell_{hT}^2 \sigma^D\rangle}{d{\cal PS}} \left/\frac{d\sigma}{d{\cal PS}}\right. ,
\qquad
\frac{d\langle \ell_{hT}^2 \sigma^D\rangle}{d{\cal PS}} 
\equiv \int d\ell_{hT}^2 \ell_{hT}^2 \frac{d\sigma^D}{d{\cal PS}d\ell_{hT}^2},
\label{eq-broadening definition}
\eea
where the phase space $d{\cal PS} = dx_Bdydz_h$ with $x_B, y, z_h$ the standard SIDIS variables, and the superscript ``$D$'' indicates the double scattering contribution. As shown in the above equation, the transverse momentum broadening is closely related to the transverse momentum $\ell_{hT}^2$-weighted differential cross section, which can be computed within a generalized factorization theorem at high-twist \cite{Luo:1994np}. 
\bef
\psfig{file=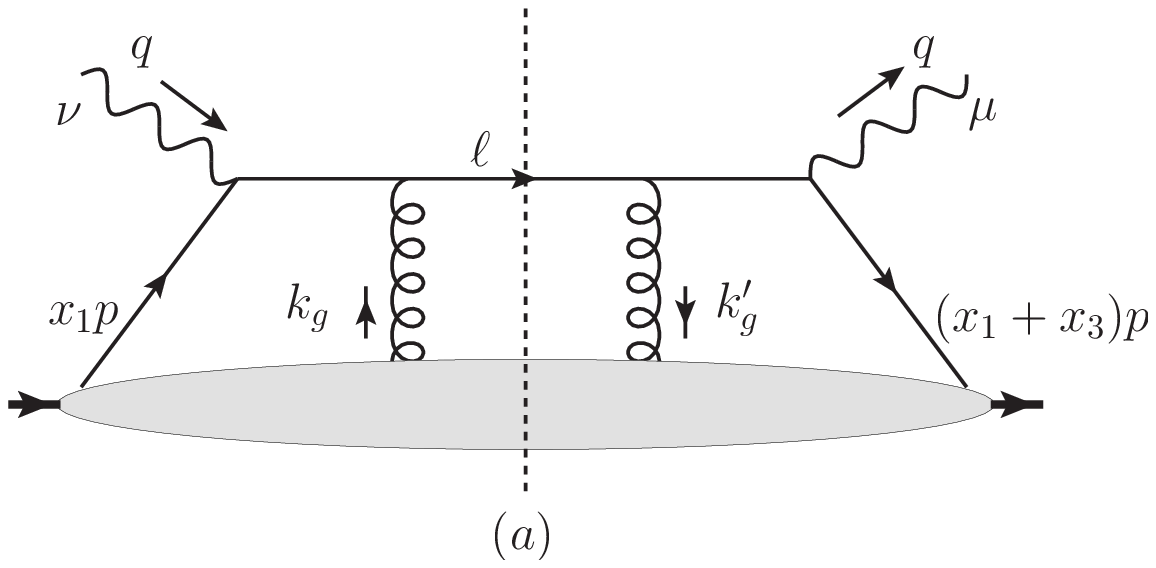, width=0.25\columnwidth}
\hskip 0.1in
\psfig{file=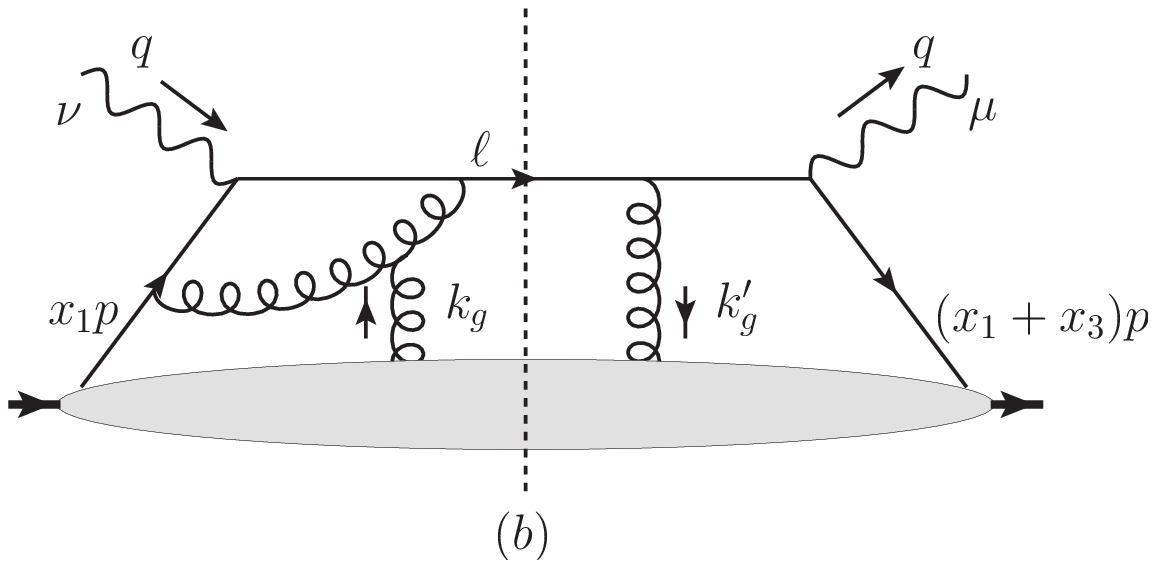, width=0.25\columnwidth}
\hskip 0.1in
\psfig{file=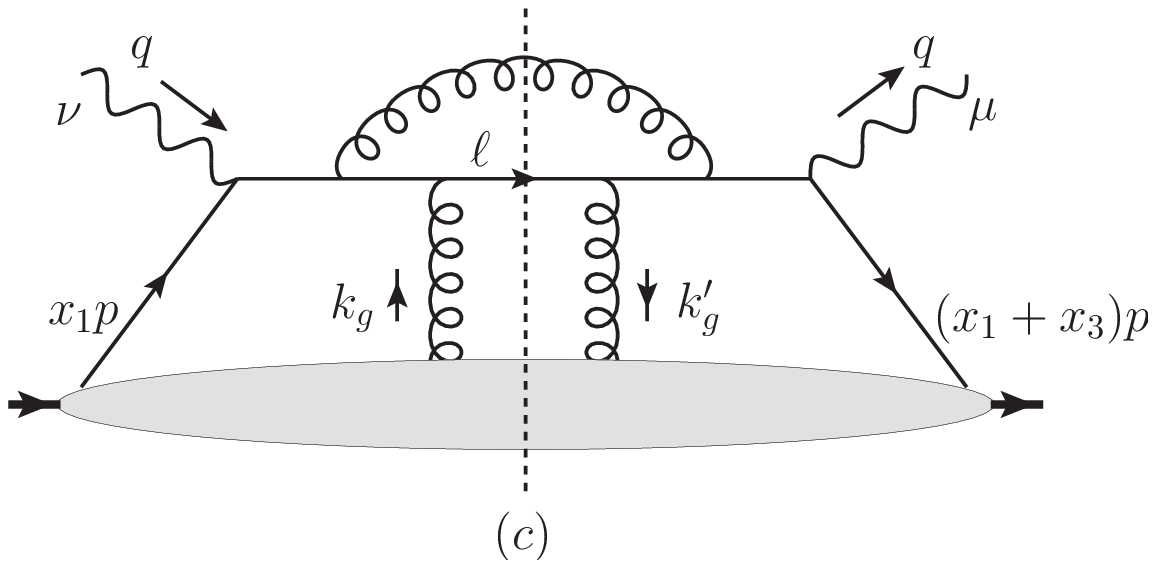, width=0.25\columnwidth}
\caption{Sample Feynman diagrams for double scattering contributions to the $\ell_{hT}^2$-weighted differential cross section from (a) leading-order  (b) NLO virtual, and (c)  NLO real processes.}
\label{diagram}
\eef
At leading order (LO), the contribution is given by Fig. \ref{diagram}(a). The calculation is straightforward, and the final result is directly proportional to the quark-gluon correlation function $T_{qg}$ as defined in \cite{Kang:2013raa},
\bea
\Delta\langle\ell_{hT}^2\rangle=\left(\frac{4\pi^2\alpha_sz_h^2}{N_c}\right)\frac{\sum_qe_q^2T_{qg}(x_B,0,0)D_{h/q}(z_h)}{\sum_qe_q^2f_{q/A}(x_B)D_{h/q}(z_h)}.
\eea
%where the quark-gluon correlation function has the following operator definitions
%\bea
%T_{qg}(x_1, x_2, x_3)
%=&\int \frac{dy^-}{2\pi} e^{ix_1p^+y^-}  \int \frac{dy_1^-dy_2^-}{4\pi} e^{ix_2p^+(y_1^- - y_2^-)}
 %e^{ix_3p^+y_2^-} \theta(y_2^-)\theta(y_1^- - y^-)
%\langle A|{\bar\psi}_q(0) \gamma^+ F_{\sigma}^+(y_2^-)F^{\sigma +}(y_1^-)\psi_q(y^-)|A\rangle.
%\label{Tqg}
%\eea
Thus the measurements of transverse momentum broadening will immediately give us the information on the twist-4 quark-gluon correlation function, and in turn the fundamental properties of the nuclear medium. 

To establish a firm QCD factorization formalism for parton multiple scattering beyond LO, we will study the NLO corrections to the transverse momentum broadening. As shown in Fig.~\ref{diagram}, NLO contribution includes both real and virtual corrections, where the perturbative calculations of such diagrams involve divergences. We will identify these divergences and understand their physical meanings. For instance, in real diagrams as shown in Fig. \ref{diagram} (c), radiative corrections reveal two types of infrared divergences from on-shell gluons. One is the so-called collinear divergences, which happens when the final-state gluon is radiated {\it collinear} to the parent quark. The other one is the so-called soft divergence, which happens when the energy of the radiated gluon approaches {\it zero}. 
To isolate these divergences and thus regularize them, we will work in $n=4-2\epsilon$ dimensions with dimensional regularization. For real corrections, we study both the partonic channels $\gamma^*+q\to q+g$ and $\gamma^*+g\to q+\bar{q}$, which involve quark-gluon and gluon-gluon scattering, respectively. The calculations are rather involved, and the final result can be expressed as follows,
\bea
\frac{d\langle\ell_{hT}^2\sigma^D\rangle^{\rm (R)}}{d{\cal PS}}=&\sigma_h\frac{\alpha_s}{2\pi}\sum_qe_q^2\int\frac{dx}{x}\int\frac{dz}{z}D_{h/q}(z)\left(\frac{4\pi\mu^2}{Q^2}\right)^{\epsilon}
\frac{1}{\Gamma(1-\epsilon)}
\Bigg\{\frac{2}{\epsilon^2}C_F\delta(1-\hat x)\delta(1-\hat z)T_{qg}(x,0,0)-\frac{1}{\epsilon}\delta(1-\hat x)
\nnu
&
\times C_F\frac{1+\hat z^2}{(1-\hat z)_+}T_{qg}(x,0,0)
-\frac{1}{\epsilon}\delta(1-\hat z)\bigg[
C_F\frac{1+\hat x^2}{(1-\hat x)_+}T_{qg}(x,0,0)
+C_A\frac{2}{(1-\hat x)_+}T_{qg}(x_B,x-x_B,0)
\nnu
&
-\frac{C_A}{2}\frac{1+\hat x}{(1-\hat x)_+}\big(T_{qg}(x,0,x_B-x)+T_{qg}(x_B,x-x_B,x-x_B)\big)
+ P_{qg}(\hat x)T_{gg}(x,0,0)\bigg]
+H^{\rm NLO-R}\otimes T\Bigg\},
\label{eq-cfinal}
\eea
where $\hat x=x_B/x$, $\hat z=z_h/z$, $P_{qq}$ and $P_{qg}$ are the usual quark-to-quark and quark-to-gluon splitting kernels in the DGLAP evolution equations, and $T_{gg}$ is the gluon-gluon correlation function \cite{Kang:2013raa}. On the other hand, the virtual correction has the following expression,
\bea
\frac{d\langle\ell_{hT}^2\sigma^D\rangle}{d{\cal PS}}^{\rm (V)}=&\sigma_h
\frac{\alpha_s}{2\pi}\int\frac{dx}{x}T_{qg}(x,0,0)\int\frac{dz}{z}D_{h/q}(z)\delta(1-{\hat x})\delta(1-{\hat z})
\left(\frac{4\pi\mu^2}{Q^2}\right)^{\epsilon}
\frac{1}{\Gamma(1-\epsilon)}C_F\left(-\frac{2}{\epsilon^2}-\frac{3}{\epsilon}-8\right).
\label{eq-virtual}
\eea

As one can see clearly in Eqs.~\eqref{eq-cfinal} and \eqref{eq-virtual}, both real and virtual corrections contain double pole ($\propto 1/\epsilon^2$), which represents the overlap of collinear and soft divergences. However, such double poles are canceled between real and virtual contributions, as required by collinear factorization. Thus we are left with only single poles ($\propto 1/\epsilon$), the collinear divergences. They represent the {\it long distance} physics, and should be a part of distribution or fragmentation functions \cite{Kang:2007nz}. It is obvious that the term associated with $\delta(1-\hat x)$ is just the collinear QCD correction to the LO quark-to-hadron fragmentation function $D_{h/q}(z_h)$,  which should be absorbed into the definition of the renormalized quark fragmentation function,
\bea
D_{h/q}(z_h,\mu_f^2)=D_{h/q}(z_h)-\frac{\alpha_s}{2\pi}\left(\frac{1}{\hat\epsilon}+\ln\frac{\mu^2}{\mu_f^2}\right)\int_{z_h}^1\frac{dz}{z}P_{qq}(\hat z)D_{h/q}(z),
\eea
where we have adopted the $\overline{\rm MS}$ scheme with $1/\hat\epsilon=1/\epsilon-\gamma_E+\ln(4\pi)$, and $\mu_f$ is the factorization scale for the fragmentation function. The factorization scale $\mu_f$-dependence leads to the well-known DGLAP evolution equation for the fragmentation function $D_{h/q}(z_h,\mu_f^2)$.

At the same time, following the same procedure of collinear factorization, one should absorb the collinear divergence associated with $\delta(1-\hat z)$ into the redefinition of the corresponding quark-gluon correlation function $T_{qg}(x_B, 0, 0)$,
\bea
T_{qg}(x_B, 0, 0, \mu_f^2)  =  T_{qg}(x_B,0,0) -\frac{\alpha_s}{2\pi}\left(\frac{1}{\hat\epsilon}+\ln\frac{\mu^2}{\mu_f^2}\right)\int_{x_B}^1\frac{dx}{x}
\Big[{\mathcal P}_{qg\to qg} \otimes T_{qg}+P_{qg}(\hat x)T_{gg}(x,0,0)\Big],
\label{eq-Tredef}
\eea 
where ${\mathcal P}_{qg\to qg} \otimes T_{qg}$ is given by
\bea
{\mathcal P}_{qg\to qg} \otimes T_{qg}  \equiv  &
P_{qq}(\hat x) T_{qg}(x, 0, 0) 
+ \frac{C_A}{2} \bigg\{ \frac{4}{(1-\hat x)_+} 
T_{qg}(x_B, x-x_B, 0) - \frac{1+\hat x}{(1-\hat x)_+}
\big[T_{qg}(x,0,x_B-x)
\nnu
&
+T_{qg}(x_B,x-x_B,x-x_B)
\big]\bigg\}.
\eea
Here we have chosen the same factorization scale $\mu_f$ as that in the fragmentation function. In principle, they do not have to be the same. After the absorption of collinear divergences into the redefinition of nonperturbative functions, we have the QCD factorized form for the $\ell_{hT}^2$-weighted cross section as:
\bea
\frac{d\langle \ell_{hT}^2\sigma^D\rangle}{d{\cal PS}} =&
\sigma_h\sum_qe_q^2\int_{x_B}^1 \frac{dx}{x}\int_{z_h}^1\frac{dz}{z}D_{h/q}(z,\mu_f^2)
\left[\delta(1-{\hat x})\delta(1-{\hat z})T_{qg}(x,0,0,\mu_f^2)
+\frac{\alpha_s}{2\pi}H^{\rm NLO} \otimes T_{qg(gg)}\right], 
\label{eq-double}
\eea
where $H^{\rm NLO} \otimes T_{qg(gg)}$ represent the finite NLO perturbative corrections to be presented in a future publication~\cite{prepare}. 

So far, our results verify the factorization of $\ell_{hT}^2$-weighted differential cross section in SIDIS from multiple scattering at twist-4 in NLO. We have also verified the factorization for the transverse momentum weighted differential cross section of Drell-Yan lepton pair production in $p+A$ collisions at twist-4 in NLO. The perturbative result contains similar collinear divergences as that in SIDIS. Besides the single pole associated with the collinear correction to beam parton distribution function from projectile proton, the same collinear correction to twist-4 quark-gluon correlation function $T_{qg}$ is found which follows the same redefinition as in SIDIS in Eq.~\eqref{eq-Tredef}. For the finite term, it is different from that in SIDIS, indicating process-dependent corrections to hard part coefficient functions. This confirms, for the first time, the collinear factorization for twist-4 observables at the NLO, and demonstrates the universality of the associated twist-4 correlation functions and in turn implies the properties of nuclear matter as probed by a propagating parton are independent of the hard processes that produce the fast partons.

\section{QCD evolution of $\hat q$}
The quark-gluon correlation function is non-perturbative which can not be calculated by pQCD. However, its perturbative change can be derived from pQCD. From Eq.~\eqref{eq-Tredef}, we obtain the evolution equation for $T_{qg}$ as,
\bea
\mu_f^2 \frac{\partial}{\partial \mu_f^2} T_{qg}(x_B,0,0,\mu_f^2)  =  \frac{\alpha_s}{2\pi} 
\int_{x_B}^1 \frac{dx}{x} \Big[{\mathcal P}_{qg\to qg} \otimes T_{qg} 
+ P_{qg}(\hat x) T_{gg}(x, 0, 0, \mu_f^2)\Big].
\label{eq-evolution}
\eea
We can further apply this equation to determine the QCD evolution of jet transport parameter $\hat q$. If one assumes the nucleus is loosely bounded, one can neglect the correlation of nucleons inside the nucleus and relate $T_{qg}$ to the jet transport parameter $\hat q$ \cite{Osborne:2002st}:
 \bea
T_{qg}(x_B,0,0,\mu_f^2) \approx \frac{N_c}{4\pi^2\alpha_{\rm s}} f_{q/A}(x_B, \mu_f^2) \int dy^- \hat {q}(\mu_f^2,y^-).
\eea
In principle, with this relation in hand, one should be able to derive the QCD evolution of $\hat q$ immediately. Unfortunately, the evolution equation in Eq.~\eqref{eq-evolution}, as it stands, is not closed (as a general feature of high-twist distributions). However, one can obtain a closed evolution equation under certain kinematic limit. For example, in the limit of large-$x_B$ ($x_B\to 1$, implying also $\hat x\to 1$), the formation time for the radiated gluon, $\tau_f=\frac{2\hat x}{\hat z(1-\hat x)}\frac{q^-}{Q^2}$, becomes much larger than nuclear size ($\tau_f\gg R_A$). Therefore, the interference between soft and hard rescatterings gives rise to destructive effect to the final contribution.  This effect is the so-called Landau-Pomeranchuk-Migdal (LPM) \cite{Migdal:1956tc} interference effect which suppresses gluon radiation with large  formation time. In this particular kinematic region, the medium effect from gluon rescattering disappears. Mathematically, this is manifest as the fact that the splitting kernel ${\mathcal P}_{qg\to qg} \otimes T_{qg}$ reduces to the vacuum one $P_{qq}(\hat x) T_{qg}(x,0,0, \mu_f^2)$. In other words, Eq.~\eqref{eq-evolution} reduces exactly to the vacuum DGLAP evolution equation, from which we find a {\it $\mu_f$-independent} $\hat q$. Such a behavior is very similar to that of the normal parton distribution function, where the scaling behavior is observed in the large-$x_B$ regime. 

To go beyond this large-$x_B$ limit, and thus study the intermediate-$x_B$ regime, we can expand the off-diagonal matrix elements $T_{qg}$ in Eq.~\eqref{eq-evolution} around $x=x_B$ and pick up the leading contribution. In this ansatz, we can derive a non-trivial evolution equation for $\hat q$. Such a evolution equation leads to not only the $\mu_f^2$-dependence, but also $x_B$-dependence of $\hat q$. This dependence essentially tells us the violation of scaling behavior of $\hat q$, again similar to that of normal parton distribution function. The $x_B$-dependence can be also translated to the energy dependence of $\hat q$, as $x_B=Q^2/2ME$ in target rest frame. The energy dependence in the intermediate-$x_B$ region is consistent with early expectations \cite{CasalderreySolana:2007sw}. The implication of such a evolution equation of $\hat q$ will be published in a future publication~\cite{prepare}.

\section{Conclusions}
We demonstrate through explicit calculation at NLO how QCD factorization holds for multiple parton scattering. In particular, we show for both SIDIS and DY processes that the transverse momentum weighted differential cross section can be factorized as the convolution of perturbative calculable hard part and non-perturbative functions, such as fragmentation function (parton distribution function) and twist-4 quark-gluon correlation function in SIDIS (DY). Through the calculation of transverse momentum broadening in SIDIS and DY, we verify the universality of the associated quark-gluon correlation function, which implies that the properties of nuclear matter as probed by a propagating parton is independent of the hard processes that create the fast partons. Our NLO analysis also identifies the QCD evolution equation for the twist-4 quark-gluon correlation function, which can be in turn applied to determine the QCD scale and jet energy dependence of the jet transport parameter. We further discuss the QCD evolution equation of $\hat q$ by looking at particular $x_B$-regions, where interesting scaling or scaling-violation behavior is found similar to that of normal parton distribution function. 
  
This work is supported by U.S. DOE under Contract No. DE-AC52-06NA25396 and No. DE-AC02-05CH11231, and within the framework of the JET Collaboration, the Major State Basic Research Development Program in China (No. 2014CB845404), and the NSFC under Grants No. 11221504 and No. 10825523, and China MOST under Grant No. 2014DFG02050, and LDRD program at LANL.

\end{document}